\documentclass[12pt]{amsart}

\usepackage{graphics}
\usepackage{amssymb}
\usepackage{amsmath}

\numberwithin{equation}{section}

\newtheorem{theorem}{Theorem}[section]
\newtheorem{assumption}{Assumption}
\newtheorem{lemma}[theorem]{Lemma}

\theoremstyle{definition}
\newtheorem{example}[theorem]{Example}
\newtheorem{definition}[theorem]{Definition}

\theoremstyle{remark}
\newtheorem{remark}[theorem]{Remark}

\newcommand{\G}{\mathbb{G}}
\newcommand{\T}{\mathbb{T}}
\newcommand{\R}{\mathbb{R}}
\newcommand{\Gcut}{{\Gamma}}
\newcommand{\Reals}{\mathbb{R}}
\newcommand{\vecpsi}{\boldsymbol{\psi}}
\newcommand{\vecphi}{\boldsymbol{\phi}}
\newcommand{\mdef}{\stackrel{\mathrm{def}}{=}}
\newcommand{\rmd}{{\mathrm d}}
\newcommand{\bbar}[1]{\overline{#1}}
\newcommand{\V}{\mathcal{V}}
\newcommand{\E}{\mathcal{E}}
\newcommand{\dirE}{\mathfrak{E}}
\newcommand{\Hs}{\mathcal{H}}
\newcommand{\Dom}{\mathcal{D}}

\usepackage{amssymb}
\usepackage{graphicx}

\begin{document}

\title{A lower bound for nodal count on discrete and metric graphs}
\author{Gregory Berkolaiko}
\address{Department of Mathematics,
  Texas A\&M University, TX 77843-3368, USA}
\email{berko@math.tamu.edu}
\thanks{This research was partially supported by NSF award number 0604859.}

\subjclass[2000]{34B45, 05C50, 15A18}

\date{}

\begin{abstract}
  We study the number of nodal domains (maximal connected regions on which a
  function has constant sign) of the eigenfunctions of Schr\"odinger operators
  on graphs.  Under certain genericity condition, we show that the number of
  nodal domains of the $n$-th eigenfunction is bounded below by $n-\ell$,
  where $\ell$ is the number of links that distinguish the graph from a tree.
  
  Our results apply to operators on both discrete (combinatorial) and metric
  (quantum) graphs. They complement already known analogues of a result by
  Courant who proved the upper bound $n$ for the number of nodal domains.
  
  To illustrate that the genericity condition is essential we show that if it
  is dropped, the nodal count can fall arbitrarily far below the number of the
  corresponding eigenfunction.
  
  In the appendix we review the proof of the case $\ell=0$ on metric trees
  which has been obtained by other authors.
\end{abstract}

\maketitle

\section{Introduction}

According to a well-know theorem by Sturm, the zeros of the $n$-th
eigenfunction of a vibrating string divide the string into $n$
``nodal intervals''.  The Courant nodal line theorem carries over one
half of Sturm's theorem to the theory of membranes: Courant proved
that the $n$-th eigenfunction cannot have more than $n$ domains.  He
also provided an example showing that no non-trivial lower bound for
the number nodal domains can be hoped for in $\Reals^d$, $d\geq2$.

But what can be said about the number of nodal domain on graphs?  Earliest
research on graphs concentrated on Laplace and Schr\"odinger operators on
discrete (combinatorial) graphs.  The functions on discrete graphs take values
on vertices of the graph and the Schr\"odinger operator is defined by
\begin{equation*}
  (H\vecpsi)_u = -\sum_{v\sim u} \psi_v + q_u \psi_u,
\end{equation*}
where the sum is taken over all vertices adjacent to the vertex $u$.

Gantmacher and Krein \cite{GantmacherKrein} proved than on a chain graph (a
tree with no branching which can be thought of as a discretization of the
interval) an analogue of Sturm's result holds: the $n$-th eigenvector changes
sign exactly $n-1$ times.  But for non-trivial graphs the situation departs
dramatically from its $\Reals^d$ analogue.  First of all, Courant's upper
bound does not always hold.  There is a correction due to multiplicity of the
$n$-th eigenvalue and the upper bound becomes\footnote{We are talking here
  about the so-called ``strong nodal domains'' --- maximal connected
  components on which the eigenfunction has a constant well-defined (i.e. not
  zero) sign} \cite{DGLS01} $n+m-1$, where $m$ is the multiplicity.  In this
paper we discuss another striking difference.  If the number of cycles of a
graph is not large, the graph behaves ``almost'' like a string:  for a typical
eigenvector, there is a lower bound on the number of nodal domains.

To be more precise, let $\ell$ be the minimal number of edges of the graph
that distinguish it from a tree (a graph with no loops).  In terms of the
number of vertices $V$ and the number of edges $E$, the number $\ell$ can be
expressed as $\ell=E-V+1$.  We show that, for a typical eigenvector, the
number of nodal domains is greater or equal to $n-\ell$.  In particular, on
trees ($\ell=0$) the nodal counting is exact: the $n$-th eigenfunction has
exactly $n$ domains.  Here by a ``typical'' eigenvector we mean an eigenvector
which corresponds to a simple eigenvalue\footnote{Thus for a ``typical''
  eigenvector the notions of ``strong'' and ``weak'' nodal domains (see
  \cite{DGLS01}) coincide} and which is not zero on any of the vertices.  This
property is stable with respect to small perturbations of the potential
$\{q_u\}$.

Another graph model on which the question of nodal domains is well-defined is
the so-called quantum or metric graphs.  These are graphs with edges
parameterized by the distance to a pre-selected start vertex.  The functions
now live on the edges of the graph and are required to satisfy matching
conditions on the vertices of the graph.  The Laplacian in this case is the
standard 1-dimensional Laplacian.  A good review of the history of quantum
graphs and some of their applications can be found in \cite{Kuc02}.

The ideas that the zeros of the eigenfunctions on the metric {\em trees\/}
behave similarly to the 1-dimensional case have been around for some time.
Al-Obeid, Pokornyi and Pryadiev \cite{AlO92,PPAO96,PP04} showed that for a
metric tree in a ``general position'' (which is roughly equivalent to our
genericity assumption~\ref{assum:simple_upto}, see Section~\ref{sec:proofs})
the number of the nodal domains of $n$-th eigenfunction is equal to $n$.  This
result was rediscovered by Schapotschnikow \cite{Schap06} who was motivated by
the recent interest towards nodal domains in the physics community
\cite{BGS02,GSW04,GSS05}.

Our result on the lower bound extends to the quantum graphs as well.
Similarly to the discrete case, we prove that even for graphs with $\ell>0$,
$n-\ell$ is a lower bound on the number of nodal domains of the $n$-th
eigenfunction.

The article is structured as follows.  In Section~\ref{sec:main_result} we
explain the models we are considering, formulate our result and review the
previous results on the nodal counting on graphs.  The case of the metric
trees has been treated before in \cite{PP04,Schap06}.  In the three remaining
cases, metric graphs with $\ell>0$, discrete trees and discrete graphs with
$\ell>0$, we believe our results to be previously unknown and in
Section~\ref{sec:proofs} we provide complete proofs.  For completeness, we
also include a sketch of the general idea behind the proofs of
\cite{PP04,Schap06} in the Appendix.  Finally, in the last subsection of
Section~\ref{sec:proofs} we show that when a graph does not satisfy our
genericity conditions, the nodal count can fall arbitrarily far below the
number of the corresponding eigenfunction.

\section{The main result}
\label{sec:main_result}

\subsection{Basic definitions from the graph theory}

Let $\G$ be a finite graph.  We will denote by $\V$ the set of its vertices
and by $\E$ the set of {\em undirected} edges of the graph.  If there exists
an edge connecting two vertices $v_1$ and $v_2$, we say that the vertices are
{\em adjacent\/} and denote it by $v_1\sim v_2$.  We will assume that $\G$ is
connected.

\begin{definition}
  \label{defn:connected}
  A graph $\G$ is {\em connected\/} if for any $v_1, v_2 \in \V$ there is a
  sequence of distinct vertices $u_1,\ldots u_n$ leading from $v_1$ to $v_2$
  ($u_1=v_1$, $u_n=v_2$ and $u_j\sim u_{j+1}$ for $j=1,\ldots n-1$).  A graph
  $\G$ is a {\em tree\/} if for any $v_1$ and $v_2$ the sequence of $u_j$
  connecting them is unique.
\end{definition}

The number of edges emanating from a vertex $v$ is called the {\em degree\/}
of $v$.  Because we only consider connected graphs, there are no vertices of
degree 0.  If a vertex $v$ has degree 1, we call it a {\em boundary\/} vertex,
otherwise we call it {\em internal}.

It will sometimes be convenient to talk about {\em directed\/} edges of the
graph.  Each non-directed edge produces two directed edges going in the
opposite directions.  These directed edges are {\em reversals\/} of each
other.  The notation for the reversal of $d$ is $\bbar{d}$; the operation of
reversal is reflexive: $\bbar{\bbar{d}} = d$.  Directed edges always come in
pairs, in other words, there are no edges that are going in one direction
only.  The set of all directed edges will be denoted by $\dirE$.  If an edge
$d$ emanates from a vertex $v$, we express it by writing $v\prec d$.

The number of vertices is denoted by $|\V|$ and the number of non-directed
edges is $|\E|$.  Correspondingly, the number of directed edges is
$|\dirE| = 2|\E|$.  

Another key definition we will need is the dimension of the cycle space of
$\G$.
\begin{definition}
  The dimension $\ell$ of the cycle space of $\G$ is the number of edges that
  have to be removed from $\E$ (leaving $\V$ as it is) to turn $\G$ into a
  connected tree.
\end{definition}

\begin{figure}[h]
  \centering
  \includegraphics{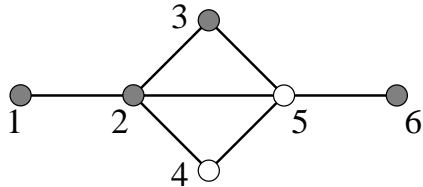}
  \caption{An example of a graph with $\ell=2$.  For example, one can cut 
    edges $(2,3)$ and $(4,5)$ to make it a tree.  If $\phi$ is positive on 
    shaded vertices and negative on white vertices, the nodal domain count 
    on the graph is 3.  On the tree obtained by deleting $(2,3)$ and $(4,5)$
    the nodal count would be 5.}
  \label{fig:graph}
\end{figure}

\begin{remark}
  An alternative characterization of $\ell$ would be the rank of the
  fundamental group of $\G$.  There is also an explicit expression for $\ell$
  in terms of the number of edges and number of vertices of the graph,
  \begin{equation}
    \label{eq:cycle_dim}
    \ell = |\E| - |\V| + 1.
  \end{equation}
  Obviously, $\ell=0$ if and only if $\G$ is a tree.
\end{remark}

\subsection{Functions on discrete graphs}

The functions on $\G$ are the functions from the vertex set $\V$ to the set of
reals, $\vecpsi: \V\to\Reals$.  We only consider finite graphs, therefore the
set of all functions $\vecpsi$ can be associated with $\Reals^{|\V|}$, where
$|\V|$ is the number of vertices of the graph.

Given a function $\vecpsi$ on $\G$, we define a {\em positive domain\/} on
$\G$ with respect to $\vecpsi$ to be a maximal connected subgraph $S$ of $\G$
such that $\vecpsi$ is positive on the vertices of $S$.  Similarly we define
the {\em negative domains}.  Then the {\em nodal domain count\/}
$\nu_{\G}(\vecpsi)$ is the total number of positive and negative domains on
$\G$ with respect to $\vecpsi$, see Fig.~\ref{fig:graph} for an example.  When
the choice of the graph is obvious, we will drop the subscript $\G$.

Our interest lies with the nodal domain counts of the eigenvectors of
(discrete) Schr\"odinger operators on graphs.  We define the Schr\"odinger
operator with the potential $q : \V\to\Reals$ by
\begin{equation}
  \label{eq:discr_schrod}
  (H\vecpsi)_u = - \sum_{v\sim u} \psi_v + q_u\psi_u.
\end{equation}
The eigenproblem for the operator $H$ is $H\vecpsi = \lambda\vecpsi$.  The
operator $H$ has $|\V|$ eigenvalues, which we number in increasing order,
\begin{displaymath}
  \lambda_1 \leq \lambda_2 \leq \ldots \leq \lambda_{|\V|}.
\end{displaymath}
This induces a numbering of the eigenvectors: $H\vecpsi^{(n)} =
\lambda_n\vecpsi^{(n)}$.  This numbering is well-defined if there are no
degeneracies in the spectrum, i.e. $\lambda_j\neq\lambda_k$ whenever $j\neq
k$.  By $\nu_{H}(\lambda_n)$ we denote the nodal domain count of the $n$-th
eigenvector $\vecpsi^{(n)}$ of an operator $H$.

\subsection{Functions on metric graphs}

A metric graph is a pair $(\G, \{L_e\})$, where $L_e$ is the length of the
edge $e\in\E$.  The lengths of the two directed edges corresponding to $e$ are
also equal to $L_e$.  In particular, $L_d = L_{\bbar{d}}$.

We would like to consider functions living on the edges of the graph.  To do
it we identify each directed edge $d$ with the interval $[0, L_d]$.  This
gives us a local variable $x_d$ on the edge which can be interpreted
geometrically as the distance from the initial vertex.  Note that if the edge
$\bar{d}$ is the reverse of the edge $d$ then $x_{\bar{d}}$ and $L_d - x_d$
refer to the same point.  Now one can define a function on an edge and,
therefore, define a function $\vecpsi$ on the whole graph as a collection of
functions $\{\psi_d\}_{d\in\dirE}$ on all edges of the graph.  To ensure that
the function is well defined we impose the condition $\psi_{d}(x_d) =
\psi_{\bar{d}}\left(L_d-x_d\right)$ for all $d\in\dirE$.  The scalar product
of two square integrable functions $\vecpsi$ and $\vecphi$ is defined as
\begin{equation}
  \label{eq:scal_prod}
  \langle \vecpsi, \vecphi\rangle \mdef  \sum_{e\in\E}\int_0^{L_e}
  \psi_e(x_e)\bbar{\phi_e(x_e)} \rmd x_e.
\end{equation}
This scalar product defines the space $L^2(\G)$.

\begin{figure}[h]
  \centering
  \includegraphics{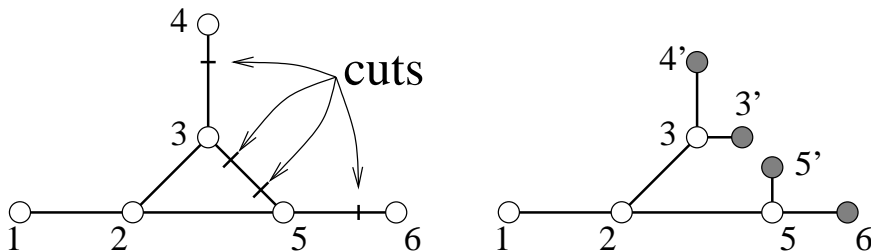}
  \caption{An example of a graph and its metric subgraph.  The shaded vertices
    are the new ones which appeared due to cuts.}
  \label{fig:metric_subgraph}
\end{figure}

To introduce the main object of our study, the nodal domains, on metric graphs
we need to define the notion of the {\em metric subgraph} of $(\G, \{L_e\})$.
\begin{definition}
  A {\em metric subgraph} of $(\G, \{L_e\})$ is a metric graph
  obtainable from $\G$ by (a) cutting some of the edges of $\G$ and thus
  introducing new boundary vertices, (b) removing some of the edges and (c)
  removing all vertices of degree 0.
\end{definition}
An example of a metric subgraph is shown on Fig.~\ref{fig:metric_subgraph}.
Now, similarly to the discrete case, we can define the nodal count for a
real-valued function $\vecphi$.

A {\em positive (negative) domain\/} with respect to a real-valued function
$\vecphi$ is a maximal connected metric subgraph on whose edges and internal
vertices $\vecphi$ is positive (corresp.\ negative).  The total number of
positive and negative domains will be called the {\em nodal count\/} of
$\vecphi$ and denoted by $\nu(\vecphi)$.

We are interested in the nodal counts of the eigenfunctions of the Laplacian
$\Delta = -\frac{\rmd^2}{\rmd x^2}$.  As its domain we take the set of {\em
  continuous\/} functions\footnote{In particular, the functions must be
  continuous across the vertices.} that belong to the Sobolev space $H^2(e)$
on each edge $e$ and satisfy the Kirchhoff condition
\begin{equation}
  \label{eq:Kirch}
  \sum_{d\succ v} \frac{\rmd}{\rmd x} \psi_d(0) = 0 \qquad
  \mbox{ for all } v\in\V.
\end{equation}
Note that the sum is taken over all directed edges that originate from the
vertex $v$ and the derivative (which depends on the direction of the edge) is
taken in the outward direction.  The Laplacian can also be defined via the
quadratic form
\begin{equation}
  \label{eq:qform_quantum}
  Q_{\Delta}[\vecpsi]
  = \sum_{e\in\E}\int_0^{L_e} |\psi_e'(x_e)|^2 \rmd x_e.
\end{equation}
The domain of this form is the Sobolev space $H^1(\G)$.

For boundary vertices condition~(\ref{eq:Kirch}) reduces to the Neumann
condition $\psi'_d(0) = 0$.  We also consider other homogeneous conditions
on the vertex $v$, of the general form
\begin{equation}
  \label{eq:gen_bc}
  \psi'_d(0) \cos \alpha_v = \psi_d(0) \sin \alpha_v,
\end{equation}
where the Neumann condition corresponds to the choice $\alpha_d=0$.  The
corresponding quadratic form will then change\footnote{if $\cos\alpha_v=0$ ---
  the Dirichlet case --- the condition $\psi(v)=0$ should instead be
  introduced directly into the domain of $Q_\Delta$} to
\begin{equation}
  \label{eq:qform_quantum_gen}
  Q_{\Delta}[\vecpsi]
  = \sum_{e\in\E}\int_0^{L_e} |\psi_e'(x_e)|^2 \rmd x_e 
  + \sum_{v:\deg(v)=1} \psi^2(v) \tan \alpha_v,
\end{equation}
where the sum is over the boundary vertices and $\psi(v)$ is the value of the
function at the vertex $v$.

Our results will also apply to Schr\"odinger operators $H = \Delta + q(x)$
with a potential $q(x)$ which is continuous\footnote{Or has finitely many
  jumps: the jumps can be thought of as ``dummy'' vertices of degree 2} on
every edge of the graph.

Schr\"odinger operator $H$, defined in the above fashion, has an infinite
discrete spectrum with no accumulation points.  As in the discrete case, we
number the eigenvalues in the increasing order.  We will denote by
$\vecpsi^{(n)}$ the eigenvector corresponding to the eigenvalue $\lambda_n$.

\subsection{Our assumptions and results}
\label{sec:assum}

Let $\lambda_n$ be the $n$-th eigenvalue of the Schr\"odinger operator $H$ on
either discrete or metric graph.  Let $\vecpsi^{(n)}$ be the corresponding
eigenfunction.  We shall make the following assumptions.

\begin{assumption}
  \label{assum:simple}
  The eigenvalue $\lambda_n$ is simple and the corresponding eigenvector
  $\vecpsi^{(n)}$ is non-zero on each vertex.
\end{assumption}

\begin{remark}
  \label{rem:genericity_of_ass}
  The properties described in the Assumption are generic and stable with
  respect to a perturbation.  Relevant perturbations include changing the
  potential $\{q_v\}$ in the discrete case and changing lengths $\{L_e\}$ in
  the metric case.  More precisely, in the finite-dimensional space of all
  potentials (corresp.\ lengths) the set $A_n$ on which
  $(\lambda_n,\vecpsi^{(n)})$ satisfy the Assumption is open and dense unless
  the graph is a circle (see \cite{Fri05}, where this question is discussed
  for metric graphs).  We also mention that on each connected component of the
  set $A_n$ the nodal count of $\vecpsi^{(n)}$ remains the same.  Indeed, on
  discrete graphs the sign of the eigenvector on each vertex must remain
  unchanged.  On metric graphs the zeros cannot pass through the vertices.
  Moreover zeros cannot undergo a bifurcation (i.e.\ appear or disappear) ---
  otherwise at the bifurcation point the eigenfunction and its derivative are
  both zero.  By uniqueness theorem for $H\vecpsi = \lambda\vecpsi$, this
  would mean that $\vecpsi$ is identically zero on the whole edge,
  contradicting the Assumption.
\end{remark}

Now we are ready to state the main theorem which applies to both discrete and
metric graphs.

\begin{theorem}
  \label{thm:main}
  Let $\lambda_n$ and $\vecpsi^{(n)}$ be the $n$-th eigenvalue and the
  corresponding eigenvector of the Schr\"odinger operator $H$ on either
  discrete or metric graph $\G$.  If $(\lambda_n,\vecpsi^{(n)})$ satisfy
  Assumption~\ref{assum:simple}, then the nodal domain count of
  $\vecpsi^{(n)}$ is bounded by
  \begin{equation}
    \label{eq:discr_bound}
    n-\ell \leq \nu(\vecpsi^{(n)}) \leq n,
  \end{equation}
  where $\ell = |\E| - |\V| + 1$ is the dimension of the cycle space of $\G$.
  In particular, when $\G$ is a tree, $\nu(\vecpsi^{(n)}) = n$.
\end{theorem}

While we state the theorem in the most complete form, we will prove only
those parts of it that we believe to be new.  The upper bound on the number of
nodal domains is a result with a long history going back to Courant
\cite{Cou23, CourantHilbert}.  The original proof for domains in $\Reals^d$
was adapted to metric graphs by Gnutzmann, Weber and Smilansky \cite{GSW04},
who used the $\Reals^d$ proof from Pleijel \cite{Ple56} who, in turn, cites
Herrmann \cite{her35} who simplified the original proof of Courant
\cite{Cou23}.

The history of the discrete version of Courant's upper bound is more
complicated.  The question was considered by Colin de Verdi\`ere \cite{CdV93},
Friedman \cite{Fri93}, Duval and Reiner \cite{DR99}, and Davies, Gladwell,
Leydold and Stadler \cite{DGLS01}.  The latter paper contains a good overview
of the history of the result and points out various shortcomings in the
preceding papers.  The point of difficulty was counting the nodal domains if
an eigenvalue is degenerate (and therefore there is an eigenvector which is
zero on some vertices).  As shown in \cite{DGLS01}, the upper bound is
$n+m-1$, where $m$ is the multiplicity of the eigenvalue.  In our case,
Assumption~\ref{assum:simple}, which is essential for the lower bound (see
Section~\ref{sec:nongeneric}), also simplifies the upper bound.

The lower bound for the nodal domains on metric trees (i.e.\ the $\ell=0$
case) was shown by Al-Obeid, Pokornyi and Pryadiev \cite{AlO92,PPAO96,PP04}
and by Schapotschnikow \cite{Schap06}.  For completeness, we give a sketch of
the proof of this case in the Appendix.

Finally, the results on the lower bound for discrete graphs (both $\ell=0$ and
$\ell>0$ cases) and for metric graphs with $\ell>0$ are new and will
be proved in this paper.

{\em Note added in proof:} It has been brought to the author's attention by
J.~Leydold that the lower bound for discrete trees has been also obtained by
B{\i}y{\i}ko{\u g}lu \cite{Biy03} as a corollary of a result of Fiedler
\cite{Fie75}.

\section{Proofs}
\label{sec:proofs}

We will apply induction on $\ell$ to deduce the statement for metric
graphs.  The proofs for the discrete case follow the same ideas but differ in
some significant detail.

First, however, we discuss an important consequence of
Remark~\ref{rem:genericity_of_ass}: it is sufficient to prove statements on
nodal counts under the following stronger Assumption.
\begin{assumption}
  \label{assum:simple_upto}
  Assumption~\ref{assum:simple} is satisfied for all eigenpairs $(\lambda_k,
  \vecpsi^{(k)})$ with $k\leq n$.
\end{assumption}
Indeed, if only Assumption~\ref{assum:simple} is satisfied but
Assumption~\ref{assum:simple_upto} is not, we can perturb the problem so
that (a) the nodal count of the $n$-th eigenfunction $\vecpsi^{(n)}$ does not
change and (b) Assumption~\ref{assum:simple} becomes satisfied for all $k\leq
n$.  Then, anything proved about the nodal domains of $\vecpsi^{(n)}$ in the
perturbed problem (which satisfies Assumption~\ref{assum:simple_upto}) will
still be valid for the unperturbed one.

In our proofs we use the classical ideas of mini-max characterization of the
eigenvalues.  Let $H$ be a self-adjoint operator with domain $\Dom$.  Assume
the spectrum of $H$ is discrete and bounded from below.  Let $Q_H[\vecpsi] =
(\vecpsi, H\vecpsi)$ be the corresponding quadratic form.  Then the
eigenvalues of $H$ can be obtained as
\begin{equation}
  \label{eq:maximin}
  \lambda_{k+1} = \max_{f_1,\ldots,f_k\in\Dom'} \  
  \min_{\vecpsi\in\Dom,\ f_j(\vecpsi)=0} 
  \frac{Q_H[\vecpsi]}{(\vecpsi,\vecpsi)},
\end{equation}
where the maximum is taken over all linear functionals over $\Dom$.  

We will need the following classical theorem (see, e.g., \cite[Chapter
VI]{CourantHilbert} or \cite[Chapter II]{Gould})
\begin{theorem}[Rayleigh's Theorem of Constraint]
  \label{thm:rayleigh}
  Let $H$ be a self-adjoint operator defined on $\Dom$.  If $H$ is restricted
  to a subdomain $\Dom_R = \{\vecpsi\in\Dom: g(\vecpsi)=0 \}$, where
  $g\in\Dom'$, then the eigenvalues $\mu_n$ of the restricted operator satisfy
  \begin{displaymath}
    \lambda_{n} \leq \mu_{n} \leq \lambda_{n+1},
  \end{displaymath}
  where $\lambda_n$ are the eigenvalues of the unrestricted operator.
\end{theorem}

\subsection{Metric graphs ($\ell>0$)}

We will derive the lower bound for graphs with cycles by cutting the cycles
and using the lower bound for trees.

\begin{proof}[Proof of Theorem~\ref{thm:main} for metric graphs ($\ell>0$)]
  We are given an eigenpair $(\lambda_n,\vecpsi^{(n)})$.  Assume that cutting
  the edges $e_1,\ldots, e_\ell$ turns the graph $\G$ into a tree.  We cut
  each of these edges at a point $x_j \in e_j$ such that $\vecpsi^{(n)}(x_j)
  \neq 0$.  We thus obtain a tree with $|\E(\G)|+\ell$ edges and
  $|\V(\G)|+2\ell$ vertices.  Denote this tree by $\T$.  There is a natural
  mapping from the functions on the graph $\G$ to the functions on the tree
  $\T$.  In particular, we can think of $\vecpsi^{(n)}$ as living on the tree.
  
  We would like to consider the same eigenproblem $H\vecpsi = \mu\vecpsi$ on
  the tree now.  The vertex conditions on the vertices common to $\T$ and $\G$
  will be inherited from the eigenproblem on $\G$.  But we need to choose the
  boundary conditions at the $2\ell$ new vertices.  Each cut-point $x_j$ gives
  rise to two vertices, which we will denote by $u_{j+}$ and $u_{j-}$.  Define
  \begin{equation*}
    a_{j+} = \frac{\frac{d}{dx}\vecpsi^{(n)}(u_{j+})}{\vecpsi^{(n)}(u_{j+})},
    \qquad
    a_{j-} = \frac{\frac{d}{dx}\vecpsi^{(n)}(u_{j-})}{\vecpsi^{(n)}(u_{j-})},
    \qquad j = 1,\ldots, \ell,
  \end{equation*}
  where the derivatives are taken in the inward direction on the corresponding
  edges of $\T$.  Since $\vecpsi^{(n)}$, as an eigenfunction, was continuously
  differentiable and $\vecpsi^{(n)}(u_{j+}) = \vecpsi^{(n)}(u_{j-})$, we have
  $a_{j+} = - a_{j-}$.  
  
  Now we set the boundary conditions on the new vertices of $\T$ to be
  \begin{equation*}
    \frac{d}{dx}\psi(u_{j+}) = a_{j+} \psi(u_{j+}), \qquad
    \frac{d}{dx}\psi(u_{j-}) = a_{j-} \psi(u_{j-}), \qquad j = 1,\ldots, \ell,
  \end{equation*}
  where the derivatives, as before, are taken inwards.  By definition of the
  coefficients $a_{j\pm}$, the function $\vecpsi^{(n)}$ satisfies the above
  boundary conditions.  It also satisfies the equation $H\vecpsi = \mu\vecpsi$
  and the vertex conditions throughout the rest of the tree.  Thus,
  $\vecpsi^{(n)}$ is also an eigenfunction on $\T$ and $\lambda_n$ is the
  corresponding eigenvalue.  If we denote the ordered eigenvalues of $\T$ by
  $\mu_k$, then $\lambda_n = \mu_m$ for some $m$.  It is important to note
  that $m$ is in general different from $n$.  We will now show that $m\geq n$.
  
  Denote by $Q_\G[\vecpsi]$ the quadratic form corresponding to the eigenvalue
  problem on $\G$; its domain we denote by $\Hs_\G$.  Similarly we define
  $Q_\T[\vecpsi]$ and $\Hs_\T$.  As we mentioned earlier, there is a natural
  embedding of $\Hs_\G$ into $\Hs_\T$.  Moreover, we can say that
  \begin{equation*}
    \Hs_\G = \left\{ \vecpsi\in\Hs_\T : \psi(u_{j+}) = \psi(u_{j-}), 
      j=1,\ldots, \ell \right\}.
  \end{equation*}
  We also note that, formally,
  \begin{equation*}
    Q_\T[\vecpsi] = Q_\G[\vecpsi] + \sum_{j=1}^\ell \left(a_{j+} \psi^2(u_{j+})
      + a_{j-} \psi^2(u_{j-}) \right).
  \end{equation*}
  If $\vecpsi\in\Hs_\G$ then $\psi(u_{j+}) = \psi(u_{j-})$ and $a_{j+} = -
  a_{j-}$ result in the cancellation of the sum on the right-hand side.  This
  means that on $\Hs_\G$, $Q_\T[\vecpsi] = Q_\G[\vecpsi]$.
  
  Now we employ the minimax formulation for the eigenvalues $\lambda_k$ on
  $\G$,
  \begin{equation*}
    \lambda_{k+1} = \max_{\phi_1,\ldots,\phi_k \in \Hs_\G}
    \min_{\stackrel{\vecpsi\in\Hs_\G}{\|\vecpsi\|=1,\ \vecpsi \perp \phi_i}} 
    Q_\G[\vecpsi] 
    = \max_{\phi_1,\ldots,\phi_k \in \Hs_\G}
    \min_{\stackrel{\vecpsi\in\Hs_\G}{\|\vecpsi\|=1,\ \vecpsi \perp \phi_i}}
    Q_\T[\vecpsi],
  \end{equation*}
  Comparing it with the corresponding formula for the eigenvalues on $\T$
  \begin{equation*}
    \mu_{k+1} = \max_{\phi_1,\ldots,\phi_k \in \Hs_\T}
    \min_{\stackrel{\vecpsi\in\Hs_\T}{\|\vecpsi\|=1,\ \vecpsi \perp \phi_i}} 
    Q_\T[\vecpsi],
  \end{equation*}
  we see that the eigenvalues $\lambda_k$ correspond to the same minimax
  problem as $\mu_k$ but with $\ell$ additional constraints $\psi(u_{j+}) =
  \psi(u_{j-})$.  By Rayleigh's theorem we conclude that $\mu_k \leq
  \lambda_k$ for any $k$.  Therefore, if $\lambda_n=\mu_m$ for some $n$ and
  $m$, they must satisfy $m\geq n$.
  
  To finish the proof we need to count the number of nodal domains on $\G$ and
  on $\T$ with respect to $\vecpsi^{(n)}$.  When we cut an edge of $\G$, we
  increase the number of nodal domains by at most one\footnote{The number of
    nodal domains might not increase at all if a nodal domain entirely covers a
    loop of $\G$}.  Therefore,
  \begin{equation*}
    \nu_\T(\vecpsi) \leq \nu_\G(\vecpsi) + \ell.
  \end{equation*}
  On the other hand, we know that the nodal counting on the tree is exact,
  and, since $\vecpsi^{(n)}$ is the $m$-th eigenvector on $\T$,
  \begin{equation*}
    \nu_\T(\vecpsi^{(n)}) = m \geq n.
  \end{equation*}
  Combining the above inequalities we obtain the desired bound
  \begin{equation*}
    \nu_\G(\vecpsi^{(n)}) \geq n - \ell.
  \end{equation*}
  
  To conclude the proof we acknowledge that we implicitly assumed that the
  tree $\T$ satisfies Assumption~\ref{assum:simple}, more precisely, that the
  eigenvalue $\mu_r$ is simple.  To justify it, we observe that, if this is
  not the case, a small perturbation in the lengths of the edges will force
  $\T$ to become generic but will not affect the properties of the
  eigenvectors of $\G$.
\end{proof}

\subsection{Discrete trees ($\ell=0$)}
\label{sec:proof_d_tree}

Take an arbitrary vertex of $\T$ and designate it as {\em root}, denoted $r$.
The tree with a root induces partial ordering on the vertices $\V$: we say
that $v_1 < v_2$ if the unique path connecting $v_1$ with $r$ passes through
$v_2$ (see Definition~\ref{defn:connected}).  We denote by $v_1 \prec v_2$ the
situation when $v_1<v_2$ and $v_1\sim v_2$.

In the above ordering the root is higher than any other vertex.  Since $\T$ is
a tree, for each vertex $v$, other than the root, there is a unique $u$ such
that $v\prec u$.  Given a non-vanishing $\vecpsi$ we introduce the new
variables $R_v = \psi_u / \psi_v$, where $v\prec u$.  Variables $R_v$ are
sometimes called {\em Riccati variables\/} \cite{MD94}.

The eigenvalue condition $H\vecpsi = \lambda\vecpsi$ can now be written as
\begin{equation}
  \label{eq:eig_cond}
  -\psi_u - \sum_{w\prec v} \psi_w + q_v\psi_v = \lambda\psi_v,
\end{equation}
and, after dividing by $\psi_v$,
\begin{equation}
  \label{eq:defn_R}
  R_v = q_v - \lambda - \sum_{w\prec v} \frac{1}{R_w}.
\end{equation}
If $v$ is the root, condition~(\ref{eq:eig_cond}) takes the form
\begin{displaymath}
  - \sum_{w\prec r} \psi_w + q_r\psi_r = \lambda\psi_r.
\end{displaymath}
Therefore, if we define
\begin{displaymath}
  R_r \equiv q_r - \lambda - \sum_{w\sim r} \frac{1}{R_w},
\end{displaymath}
then the zeros of $R_r$ in terms of $\lambda$ are the eigenvalues of $H$.
Whenever $R_r(\lambda)=0$, the values of $R_v$, $v\neq r$, uniquely specify
the corresponding eigenvector $\vecpsi$ of $H$, and vice versa.

Equation (\ref{eq:defn_R}) provides a recursive algorithm for calculating
$R_v$, in order of increasing $v$.  Thus one gets a closed formula for $R_v$
in terms of $q_u$, $u\leq v$ and $\lambda$. This is best illustrated by an
example.

\begin{figure}[h]
  \centering
  \includegraphics{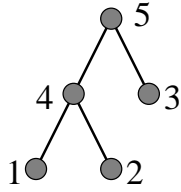}
  \caption{An example of a tree graph with $5$ being the root vertex.}
  \label{fig:tree}
\end{figure}

\begin{example}
  For the tree shown in Fig.~\ref{fig:tree} the eigenvalue condition in terms
  of Riccati variables reads
  \begin{eqnarray}
    \label{eq:R1}
    R_1 &=& q_1 - \lambda,\\
    \label{eq:R2}
    R_2 &=& q_2 - \lambda,\\
    \label{eq:R3}
    R_3 &=& q_3 - \lambda,\\
    \label{eq:R4}
    R_4 &=& q_4 - \lambda - \frac1{R_1} - \frac1{R_2},\\
    \label{eq:R5}
    0 &=& q_5 - \lambda - \frac1{R_3} - \frac1{R_4} \equiv R_5.
  \end{eqnarray}
  By substituting lines (\ref{eq:R1}) and (\ref{eq:R2}) into equation
  (\ref{eq:R4}), and then lines (\ref{eq:R3}) and (\ref{eq:R4}) into equation
  (\ref{eq:R5}), one obtains an eigenvalue condition for $H$. 
\end{example}

Denote by $P_v$ the set of all poles of $R_v$ with respect to $\lambda$ and by
$Z_v$ the set of all zeros of $R_v$; these sets are finite.  We define
$N_v^{<}$ to be the number of negative values among $R_u$ with $u<v$; we
similarly define $N_v^{\leq}$:
\begin{equation}
  \label{eq:negatives_defn}
  N_v^< = \Big|\{u<v: R_u < 0\}\Big|, \qquad 
  N_v^\leq = \Big|\{u \leq v: R_u < 0\}\Big|.
\end{equation}
The above numbers are not defined whenever one of $R_u$ has a zero or a pole.
The following lemma, listing properties of the Riccati variables, their poles
and zeros, amounts to the proof of Theorem~\ref{thm:main} when $\G$ is a tree
and $q$ is generic.

\begin{lemma}
  \label{lem:polesNzeros}
  Assume that, for each $v$, the sets $Z_w$ with $w\prec v$ are pairwise
  disjoint for all $v$.  Then
  \begin{enumerate}
  \item \label{itm:poles_n_zeros} 
    $P_v = \bigcup_{w\prec v} Z_w$
  \item \label{itm:decay} For every $p\in P_v$, $lim_{\lambda\to p-0} R_v =
    -\infty$ and $lim_{\lambda\to p+0} R_v = +\infty$.  Also,
    $lim_{\lambda\to-\infty} R_v = +\infty$ and $lim_{\lambda\to\infty} R_v =
    -\infty$.  Outside the poles, $R_v$ is continuous and monotonically
    decreasing as a function of $\lambda$.
  \item \label{itm:interlacing}
    There is exactly one zero of $R_v$ strictly between each pair of
    consecutive points from the set $\{-\infty\} \cup \{\infty\} \cup P_v$.
  \item \label{itm:Nleq} Between each pair of consecutive points from
    $\{-\infty\} \cup \{\infty\} \cup Z_v$, the number $N_v^\leq$ (where
    defined) remains constant.  When a zero of $R_v$ is crossed, $N_v^\leq$
    increases by one.
  \item \label{itm:Nless} Between each pair of consecutive points from
    $\{-\infty\} \cup \{\infty\} \cup P_v$, the number $N_v^<$ (where defined)
    remains constant.  When a pole of $R_v$ is crossed, $N_v^<$ increases by
    one.
  \item \label{itm:nodal_count}
    When $\lambda=\lambda_n$ is an eigenvalue of $H$, the number of the
    nodal domains of $\vecpsi^{(n)}$ is given by
    \begin{equation}
      \label{eq:nu_via_R}
      \nu(\lambda_n) = N_r^< + 1.
    \end{equation}
  \end{enumerate}
\end{lemma}

\begin{proof}
  Part \ref{itm:poles_n_zeros} follows directly from equation
  (\ref{eq:defn_R}).
  
  Part \ref{itm:decay} follows from (\ref{eq:defn_R}) by induction over
  increasing $v$.
  
  Part \ref{itm:interlacing} follows from part \ref{itm:decay}: between each
  pair of consecutive points from $\{-\infty\} \cup \{\infty\} \cup P_v$, the
  function $R_v$ decreases from $+\infty$ to $-\infty$.
  
  Parts \ref{itm:Nleq} and \ref{itm:Nless} are linked together in an induction
  over increasing $v$.  The induction is initialized by $N_v^\leq$ for minimal
  $v$ (i.e.\ there is no $w$ with $w<v$).  In this case, $R_v = q_v -
  \lambda$, therefore $N_v^\leq = 0$ to the left of $\lambda=q_v$ and
  $N_v^\leq = 1$ to the right of $\lambda=q_v$.
  
  The inductive step starts with part \ref{itm:Nless}.  For a vertex $v$, let
  both statements be verified for all $w$, $w<v$.  The statement for $N_v^<$
  is obtained immediately from the duality between the zeros and the poles
  (part \ref{itm:poles_n_zeros}).  Note that the assumption of the lemma
  implies that only one of $N_w^\leq$ with $w\prec v$ can increase when
  $\lambda$ crosses a pole of $R_v$.

  To obtain the statement for $N_v^\leq$ consider two consequent poles and two
  consequent zeros of $R_v$, interlacing as follows
  \begin{displaymath}
    p_1 < z_1 < p_2 < z_2.
  \end{displaymath}
  Then $R_v$ is positive for $\lambda\in(p_1, z_1)$ (by part \ref{itm:decay}),
  therefore, on this interval $N_v^\leq = N_v^<$.  When $z_1$ is crossed,
  $N_v^\leq$ increases by one since $R_v$ becomes negative: $N_v^\leq = N_v^<
  + 1 \equiv C$.  On the other hand, when $p_2$ is crossed, $N_v^\leq$ and
  $N_v^<$ become equal again since $R_v>0$.  However, $N_v^<$ has increased by
  one (by the induction hypothesis) and therefore $N_v^\leq$ is still equal to
  $C$.  The above is obviously valid even if $p_1=-\infty$ or/and
  $z_2=+\infty$.
  
  Finally, to show part \ref{itm:nodal_count} we observe that $N_r^<$ is the
  number of negative Riccati variables throughout the tree.  If $R_v<0$ then
  the signs of $\psi_v$ and $\psi_u$ (where $u$ is the unique vertex
  satisfying $v\prec u$) are different, i.e.\ the edge $(u,v)$ is a boundary
  between a positive and a negative domain.  Removing all boundary edges
  separates the tree into subtrees corresponding to the positive/negative
  domains.  But removing $N_r^<$ edges from a tree breaks it into $N_r^<+1$
  disconnected components, therefore the number of domains on a tree is equal
  to $N_r^<+1$.
\end{proof}

\begin{proof}[Proof of Theorem~\ref{thm:main} for discrete trees ($\ell=0$)]
  The condition of Lemma~\ref{lem:polesNzeros} is satisfied due to the
  genericity assumption.  Indeed, if there are $v$, $w_1$ and $w_2$ such that
  $w_1\prec v$, $w_2\prec v$ and $\lambda \in Z_{w_1}\cap Z_{w_2}$ then one
  can construct an eigenvector with eigenvalue $\lambda$ and with $\psi_v=0$.
  
  Since the sets $Z_v$ are finite, $N_r^<$ must become zero when
  $\lambda\to-\infty$.  Consequently, $N_r^<$ is zero between $-\infty$ and the
  first pole of $R_r$.  Denote by $p_n$ the $n$-th pole of $R_r$.  By
  part~\ref{itm:interlacing} of the lemma, the first eigenvalue $\lambda_1$ of
  $H$ lies in the interval $(\infty, p_1)$, on which $N_r^<$ is zero.  By
  (\ref{eq:nu_via_R}) we thus have $\nu(\lambda_1)=1$.  Further, $\lambda_2$
  lies in the interval $(p_1,p_2)$.  By part \ref{itm:Nless} of the lemma,
  $N_r^<=1$ on this interval, giving $\nu(\lambda_2)=2$.  Equality for other
  $\lambda_n$ follows similarly.
\end{proof}

\subsection{Discrete graphs ($\ell>0$)}

In this case $H$ is a matrix and the quadratic form is 
\begin{equation}
  \label{eq:qform_def}
  Q_\G[\vecpsi] = Q_H[\vecpsi] = \sum_{j,k=1}^{|\V|} H_{jk}\psi_j\psi_k,
\end{equation}
where
\begin{displaymath}
  H_{jk} = 
  \begin{cases}
    -1, & j\sim k,\\
    q_j, & j=k,\\
    0, & \mbox{otherwise}.
  \end{cases}
\end{displaymath}

\begin{proof}[Proof of Theorem~\ref{thm:main} for discrete graphs ($\ell>0$)]
  We will prove the result by induction.  The initial inductive step $\ell=0$ is
  already proven in Section~\ref{sec:proof_d_tree}.
  
  Assume, without loss of generality, that we can delete the edge $(1,2)$ of
  the graph $\G$ without disconnecting it.  We will denote thus obtained graph
  by $\Gcut$.  Note that $\V(\G) = \V(\Gcut)$.  Let $\vecphi$ be an
  eigenvector of $H_\G$ with eigenvalue $\lambda_n$.  We would like to prove
  that $\nu_\G(\vecphi) \geq n-l$.

  Set $\alpha=\phi_2/\phi_1$ and define the potential $p$ on $\Gcut$ by
  \begin{displaymath}
    p_j = 
    \begin{cases}
      q_1 - \alpha, & j=1,\\
      q_2 - 1/\alpha, & j=2,\\
      q_j, & j\neq 1,2.
    \end{cases}
  \end{displaymath}
  With the aid of potential $p$ we define the operator $H_\Gcut$ in the usual
  way, see equation (\ref{eq:discr_schrod}).  It is easy to see that, due to
  our choice of potential $p$, the vector $\vecphi$ is an eigenvector of
  $H_\Gcut$.  For example,
  \begin{displaymath}
    \left(H_\Gcut \vecphi\right)_1 = -\sum_{j\sim 1}\phi_j + (q_1-\alpha)\phi_1
    = -\sum_{j\sim 1}\phi_j - \phi_2 + q_1 \phi_1 
    = \left(H_\G \phi\right)_1 = \lambda_n \phi_1,
  \end{displaymath}
  where the adjacency is taken with respect to the graph $\Gcut$.
  
  The eigenvalue corresponding to $\vecphi$ remains unchanged.  However, in the
  spectrum $\{\mu_j\}_{j=1}^{|\V|}$ of $H_\Gcut$, this eigenvalue may occupy a
  position other than the $n$-th.  We denote by $m$ the new position of
  $\lambda_n$: $\mu_m=\lambda_n$.

  Now consider the quadratic form associated with $H_\Gcut$.  Consulting
  (\ref{eq:qform_def}) we conclude
  \begin{equation}
    \label{eq:qform_cut}
    Q_\Gcut[\vecpsi] = Q_\G[\vecpsi] + 2\psi_1\psi_2 - \alpha \psi_1^2 
    - \alpha^{-1}\psi_2^2.
  \end{equation}

  Consider first the case $\alpha>0$.  We write $Q_\Gcut[\vecpsi]$ in the form
  \begin{displaymath}
    Q_\Gcut[\vecpsi] = Q_\G[\vecpsi] - (\alpha^{1/2}\psi_1 - \alpha^{-1/2}\psi_2)^2
    \leq Q_\G[\vecpsi].
  \end{displaymath}
  From here and equation~(\ref{eq:maximin}) we immediately conclude that
  $\mu_j\leq \lambda_j$.  Therefore, $\mu_m=\lambda_n$ implies $m\geq n$.
  
  From the inductive hypothesis we know that $\nu_\Gcut[\vecphi] \geq m -
  (l-1)$.  But the number of nodal domains of $\vecphi$ with respect to
  $\Gcut$ is either the same or one more than the number with respect to $\G$:
  $\alpha>0$, therefore $\phi_1$ and $\phi_2$ are of the same sign and we may
  have cut one domain in two by deleting the edge $(1,2)$.  In particular,
  $\nu_\Gcut[\vecphi] \leq \nu_\G[\vecphi] + 1$.  Eliminating
  $\nu_\Gcut[\vecphi]$, we obtain $\nu_\G[\vecphi] + 1 \geq m - (l-1)$,
  which is the sought conclusion.

  In the case $\alpha<0$ the quadratic form on $\Gcut$ can be written as
  \begin{equation}
    \label{qform_cut_beta}
    Q_\Gcut[\vecpsi] = Q_\G[\vecpsi] 
    + (\beta^{1/2}\psi_1 + \beta^{-1/2}\psi_2)^2,
  \end{equation}
  where $\beta = -\alpha$.  Consider the subspace
  \begin{displaymath}
    \Dom_R = \{\vecpsi\in\Reals^{|\V|}:  
    \beta^{1/2}\psi_1 + \beta^{-1/2}\psi_2 = 0\}.
  \end{displaymath}
  The restrictions of $H_\G$ and $H_\Gcut$ to this subspace coincide, as can
  be seen from (\ref{qform_cut_beta}).  Therefore we can apply
  Theorem~\ref{thm:rayleigh} twice, obtaining
  \begin{displaymath}
    \lambda_{j-1} \leq \rho_{j-1} \leq \lambda_j \qquad
    \mu_{j-1} \leq \rho_{j-1} \leq \mu_j,
  \end{displaymath}
  where $\rho_j$ are the eigenvalues of the restricted operator.  In
  particular, we conclude that $\mu_{j-1}\leq\lambda_j$.  Since
  $\lambda$-spectrum is non-degenerate, $\mu_{j-1}<\lambda_{j+1}$, therefore
  $\mu_m=\lambda_n$ implies $m \geq n-1$.
  
  On the other hand, the number of nodal domains with respect to $\Gcut$ is
  the same as with respect to $\G$: since $\alpha=\phi_2/\phi_1<0$, we have
  cut an edge {\em between\/} two domains.  Using the inductive hypothesis we
  conclude that
  \begin{displaymath}
    \nu_\G(\vecphi) = \nu_\Gcut(\vecphi) \geq m - (l-1) \geq n - 1 - (l-1) 
    = n-l.
  \end{displaymath}
  
  We finish the proof with a remark similar to the final statement of the
  proof for metric graphs.  If the new graph $\Gcut$ happens not to satisfy
  Assumption~\ref{assum:simple}, a small perturbation in $q$ will force
  $\Gcut$ to become generic but will not affect the properties of the
  eigenvectors of $\G$.
\end{proof}

\subsection{Low nodal count in a non-generic case}
\label{sec:nongeneric}

In this section we show that the genericity assumption
(Assumption~\ref{assum:simple}) is essential for the existence of the lower
bound.  We shall construct an example in which the assumption is violated and
the nodal count becomes very low.  The construction is based on the fact that
an eigenfunction of a graph (as opposed to a connected domain in $\Reals^d$)
may be identically zero on a large set.

We consider a metric star graph, which is a tree with $N$ edges all connected
to a single vertex.  For Dirichlet boundary conditions one can show \cite{KS99}
that $k^2$ is an eigenvalue of the graph if
\begin{equation}
  \label{eq:star_tan}
  \sum_{j=1}^N \cot kL_j = 0.
\end{equation}
To obtain all eigenvalues of the star graph, one needs to add to the solutions
of (\ref{eq:star_tan}) the points which are ``multiple'' poles of the
left-hand side of (\ref{eq:star_tan}).  More precisely, if a given $k$ is a
pole for $m$ cotangents at the same time, then $k^2$ is an eigenvalue of
multiplicity $m-1$.  Those eigenvalues that are not poles (but zeros) of the
left-hand side of (\ref{eq:star_tan}) interlace the poles: between each pair
of consecutive poles (coming from different cotangents) there is exactly one
zero.

Now we choose the lengths $L_j$ to exploit the above features.  Let $L_1=1$,
$L_2=1/m$ for some $m\in\mathbb{N}$, and the remaining lengths be irrational
pairwise incommensurate numbers slightly greater than 1.  By construction,
$k=m\pi$ is a pole for $\cot(kL_1)$ and $\cot(kL_2)$.  The corresponding
eigenfunction is a sine-wave on the edges $1$ and $2$ and is zero on the other
edges.  It is easy to see that it has $m+1$ nodal domains.  On the other hand,
counting the poles of (\ref{eq:star_tan}), one can deduce that there are
$(m-1)(N-1)+1$ eigenvalues preceding $(m\pi)^2$.  Thus, we have constructed an
eigenfunction which is very high in the spectrum but has low number of nodal
domains.

A similar construction is possible for discrete graphs as well.

\section*{Acknowledgment}

The result of the present article came about because of two factors.  The
first was the request by Uzy Smilansky that the author give a talk on the
results of \cite{Schap06} at the workshop ``Nodal Week 2006'' at Weizmann
Institute of Science.  The second was the discussion the author had with Rami
Band on his proof that the nodal count resolves isospectrality of two graphs,
one with $\ell=0$ and the other with $\ell=1$ (now a part of
\cite{BanShaSmi06}).  Rami showed that the nodal count of the latter graph is
$\nu(\vecpsi^{(n)})=n-1$ or $n$ with equal frequency.  His result lead the
author to conjecture that for the graphs close to trees the nodal count of the
$n$-th eigenstate does not stray far from $n$.  The author is also grateful to
Uzy Smilansky and Rami Band for patiently listening to the reports on the
progress made in the proof of the conjecture and carefully checking the draft
of the manuscript.

The author is indebted to Leonid Friedlander for his explanations of the
results and techniques of \cite{Fri05}.  The author is also grateful to Tsvi
Tlusty for pointing out reference \cite{DGLS01}, to Vsevolod Chernyshev
for pointing out \cite{PPAO96,PP04}, to Vladimir Pryadiev for pointing out
\cite{AlO92} and to Philipp Schapotschnikow for several useful comments.

Most of the work was done during the author's visit to the Department of
Physics of Complex Systems, Weizmann Institute of Science, Israel.

\appendix

\section{Ideas behind the proof for metric trees ($\ell=0$)}

In this section we give an informal overview of the proof of
(\ref{eq:discr_bound}) on a metric tree ($\ell=0$).  For detailed and
rigorous proofs we refer the reader to \cite{PPAO96,PP04,Schap06}.

\begin{figure}[h]
  \centering
  \includegraphics{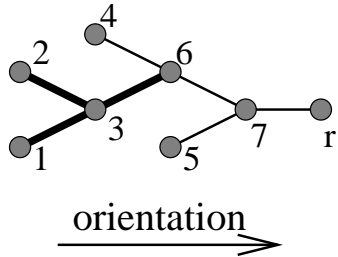}
  \caption{An example of a tree with root $r$.  If $v$ is vertex 6 then it has
    two incoming subtrees, one consisting of edges $(1,3)$, $(2,3)$ and
    $(3,6)$ (highlighted in thicker lines) and the other consisting of only
    one edge $(4,6)$.  The outgoing edge of $v$ is the edge $(6,7)$.}
  \label{fig:subtree}
\end{figure}

Let $(\lambda_n, \vecpsi^{(n)})$ be an eigenpair for a tree $\T$ satisfying
Assumption~\ref{assum:simple_upto}.  Choose an arbitrary boundary vertex of
the tree $\T$ and call it the {\em root} $r$.  We can now orient all edges of
the tree {\em towards\/} the root (well-defined because it is a tree) and will
be taking derivatives in this direction.  For each non-root vertex $v$ there
is only one adjacent edge that is directed away from it.  We call it the {\em
  outgoing\/} edge of the vertex $v$.  The other adjacent edges are
correspondingly {\em incoming}.  An {\em incoming subtree\/} of vertex $v$ is
defined recursively as the union of an incoming edge $(u,v)$ with all incoming
subtrees of the vertex $u$, see Fig.~\ref{fig:subtree}.

If we drop the boundary condition at the root, then for any $\lambda\leq
\lambda_n$ there is a solution $\vecphi(\lambda, x)$ which solves the equation
$H\vecphi = \lambda\vecphi$ and satisfies all remaining vertex conditions.
This solution is unique up to a multiplicative constant.

The function $\vecphi$ can be constructed recursively.  We fix $\lambda$ and
initialize the recursion by solving the equation $H\vecphi = \lambda\vecphi$
on the outgoing edge of each non-root boundary vertex and imposing the
boundary condition corresponding to this vertex.

Now let $v$ be a vertex such that the equation is solved on each incoming
subtree $\T^{v}_j$.  We denote these solutions (which are defined up to
a multiplicative constant) by $C_j\phi_j(x)$.  We would like to match
these solutions and to extend them to the outgoing edge of $v$.

Denoting the solution of the outgoing edge by $\phi_v(x)$ we write out the
matching conditions at the vertex $v$,
\begin{align*}
  \phi_v(v) &= C_1\phi_1(v) = C_2\phi_2(v) = \ldots\\
  \phi_v'(v) &= C_1\phi_1'(v) + C_2\phi_2'(v) + \ldots
\end{align*}

Suppose that all of the functions $\phi_j(x)$ assume non-zero values on the
vertex $v$.  Then the condition on $\phi_v(x)$ takes the form
\begin{equation*}
  \phi_v'(v) = \phi_v(v) \left( \frac{\phi_1'(v)}{\phi_1(v)} +
  \frac{\phi_2'(v)}{\phi_2(v)} + \ldots \right).
\end{equation*}
It is now clear that $\phi_v$, as a solution of $H\vecphi = \lambda\vecphi$
satisfying this condition, is also defined up to a multiplicative constant,
$C_v$.  The continuity condition now fixes the constants $C_j$ to be $C_v
\phi_v(v)/\phi_j(v)$.  Thus we obtain the solution on the union of subtrees
$\T^{v}_j$ and the outgoing edge of $v$.  This union is in turn an incoming
subtree for another vertex (or the root).

In the case when one of $\phi_j(x)$ is zero on the vertex $v$ (without loss of
generality we take $\phi_1(v) = 0$), the condition on $\phi_v$ takes the form
$\phi_v(v) = 0$.  The solution $\phi_v$ is again defined up to a
multiplicative constant $C_v$.  The values of the other constants are now
given by $C_1 = C_v \phi_v'(v) / \phi_1'(v)$ and $C_j=0$ when $j>1$.  Again
the solution on the union of subtrees $\T^{v}_j$ and the outgoing edge of $v$
is obtained up to a constant.

Finally, if more than one of $\phi_j(x)$ is zero on the vertex $v$ (without
loss of generality, $\phi_1(v) = \phi_2(v) = 0$), one can take $C_j=0$ for all
$j>2$, find non-zero $C_1$ and $C_2$ such that $C_1\phi'_1(v) + C_2 \phi'_2(v)
= 0$ and extend the function by zero on the rest of the tree.  This function
will satisfy the Kirchhoff condition at $v$ and also all other vertex
conditions.  Thus it is an eigenfunction and, moreover, it is equal to zero at
an inner vertex.  This contradicts our assumptions.

We have now constructed a function $\vecphi(\lambda,x)$ which coincides with
the eigenfunction of the tree whenever it satisfies the boundary condition at
the root.  
To count the nodal domains we need to understand the behavior of zeros of
$\vecphi$ as we change $\lambda$.  In order to do that we consider the
function\footnote{sometimes called the Weyl-Titchmarsh function or
  Dirichlet-to-Neumann map} $R(\lambda,x) = \vecphi'(\lambda,x) /
\vecphi(\lambda,x)$ where the derivative is taken with respect to $x$ in the
direction towards the root.  If $x$ is a zero of $\vecphi$, it becomes a pole
of $R(\lambda,x)$.  From the definition of $R(\lambda,x)$ we see that
$R(\lambda, x-0)=-\infty$ and $R(\lambda, x+0)=\infty$.  Differentiating
$R(\lambda,x)$ with respect to $x$ and using the equation $-\vecphi'' +
q(x)\vecphi = \lambda\vecphi$, we see that $R(\lambda,x)$ satisfies
\begin{equation*}
  \frac{d}{dx}R = q(x) - \lambda - R^2,
\end{equation*}
a Riccati-type equation.  Conditions (\ref{eq:gen_bc}) on the boundary
vertices in terms of $R(\lambda,x)$ take the form $R(\lambda,v) =
\tan(\alpha_v) \in \R^1\cup \{\infty\}$.  The matching conditions on the
internal vertices imply that the value of $R(\lambda,v)$ on the outgoing edge
is equal to the sum of the values of $R(\lambda,v)$ on the incoming edges (in
general, $R$ is not continuous on internal vertices).

Now let $\lambda_2 > \lambda_1$ and $R(\lambda_2, x_0) = R(\lambda_1, x_0)$.
Then $R'(\lambda_2, x) < R'(\lambda_1, x)$ and therefore, on some interval
$(x_0, x_0+\epsilon)$, we have $R(\lambda_2, x) < R(\lambda_1, x)$.  Moreover,
once $R(\lambda_2,x) \leq R(\lambda_1, x)$, we have $R(\lambda_2,y) \leq
R(\lambda_1, y)$ for all $y>x$ provided both functions do not have poles on
$[x,y]$.  This can be seen by assuming the contrary and considering the
point $z\in[x,y]$ where $R(\lambda_2, z) = R(\lambda_1, z)$.

Using these properties one can conclude that for each fixed $x_0$, the value
$R(\lambda, x_0)$ is decreasing as a function of $\lambda$ between the pairs
of consecutive poles.  A direct consequence of this is that the poles of $R$
move in the ``negative'' direction as the parameter $\lambda$ is increased.
The zeros of $\vecphi$, therefore, move in the direction from the root to the
leaves.  Since $q(x)$ is continuous, zeros of $\vecphi$ cannot bifurcate on
the edges, see Remark~\ref{rem:genericity_of_ass} in Section~\ref{sec:assum}.

To see that the zeros of $\vecphi$ do not split when passing through the
vertices, assume the contrary and consider the reverse picture: $\lambda$ is
decreasing.  There are at least two subtrees with zeros of $\vecphi$
approaching the same vertex $v$ as $\lambda$ approaches some critical value
from above.  At this critical value we thus have two subtrees on which
$\vecphi$ has zero at $v$.  But earlier we concluded that this situation
contradicts our genericity assumption.

To summarize, as $\lambda$ is increased, new zeros appear at the root and
move towards the leaves of the tree.  The zeros already in the tree do not
disappear or increase in number.  Now suppose $\lambda_k$ is an eigenvalue and
thus $R(\lambda_k,r) = \tan(\alpha_r)$.  As we increase $\lambda$ the value of
$R(\lambda,r)$ {\em decreases} to $-\infty$, jumps to $+\infty$ (when a new
zero enters the tree) and then increases to $\tan(\alpha_r)$ again.  Thus
between each pair of eigenvalues exactly one new zero enters the tree.  And,
on a tree, the number of nodal domains is equal to the number of internal
zeros plus one.

\def\cprime{$'$}
\providecommand{\bysame}{\leavevmode\hbox to3em{\hrulefill}\thinspace}
\providecommand{\href}[2]{#2}

\end{document}